\documentclass[aps,prb,reprint,groupedaddress,notitlepage,longbibliography,superscriptaddress]{revtex4-1}
\usepackage{amsmath}
\usepackage{graphicx}
\usepackage{natbib}
\usepackage{comment}
\usepackage{color}
\usepackage{hhline}
\usepackage{chngcntr}
\usepackage{multirow}
\usepackage{array}
\usepackage{textcomp}

\begin{document}
\title{Refraction efficiency of Huygens' and bianisotropic terahertz metasurfaces}
\author{Michael A. Cole}
\affiliation{Nonlinear Physics Centre, Research School of Physics and Engineering, The Australian National University, Canberra ACT 2601, Australia}
\email{michael.cole@anu.edu.au}	
\author{Aristeidis Lamprianidis}
\affiliation{Nonlinear Physics Centre, Research School of Physics and Engineering, The Australian National University, Canberra ACT 2601, Australia}
\affiliation{Department of Mathematics and Applied Mathematics, University of Crete, 70013 Heraklion, Crete, Greece}
\author{Ilya V. Shadrivov}
\affiliation{Nonlinear Physics Centre, Research School of Physics and Engineering, The Australian National University, Canberra ACT 2601, Australia}
\author{David A. Powell}
\affiliation{School of Engineering and Information Technology, University of New South Wales, Canberra, ACT 2610, Australia}
\affiliation{Nonlinear Physics Centre, Research School of Physics and Engineering, The Australian National University, Canberra ACT 2601, Australia}

\begin{abstract}
Metasurfaces are an enabling technology for complex wave manipulation functions, including in the terahertz frequency range, where they are expected to advance security, imaging, sensing, and communications technology. For operation in transmission, Huygens' metasurfaces are commonly used, since their good impedance match to the surrounding media minimizes reflections and maximizes transmission. Recent theoretical work has shown that Huygens' metasurfaces are non-optimal, particularly for large angles of refraction, and that to eliminate reflections and spurious diffracted beams it is necessary to use a bianisotropic metasurface. However, it remains to be demonstrated how significant the efficiency improvement is when using bianisotropic metasurfaces, considering all the non-ideal features that arise when implementing the metasurface design with real meta-atoms. 
Here we compare concrete terahertz metasurface designs based on the Huygens' and Omega-type bianisotropic approaches, demonstrating anomalous refraction angles for 55$^\circ$, and 70$^\circ$. We show that for the lower angle of 55$^\circ$, there is no significant improvement when using the bianisotropic design, whereas for refraction at 70$^\circ$ the bianisotropic design shows much higher efficiency and fidelity of refraction into the designed direction. We also demonstrate the strong perturbations caused by near-field interaction, both between and within cells, which we compensate using numerical optimization.
\end{abstract}

\maketitle

\section{Introduction}
Terahertz (THz) radiation is non-ionizing, it readily passes through many common materials, and it can be used to detect the unique spectral signatures of many chemicals \cite{grbovic2013,with2014,cheng2015,wilmink2011,zandonella2003,choi2014,witha2009}. These properties have the potential for significant advances in current technologies, such as imaging, security, biomedical analysis, and communications \cite{grbovic2013,with2014,cheng2015,wilmink2011,hafez2016, tonouchi2007, jepsen2011}.  Metasurfaces are expected to be a key platform in many of these applications, due to the flexibility of wave manipulation which they enable.
An important advance was the introduction of the Huygens' metasurface (HMS) with matched electric and magnetic dipole responses, which minimizes unwanted reflections \cite{pfeiffer2013}. Such designs have been utilized in application including perfect transmission, reflection and absorption, as well as, holography and beam steering \cite{moitra2014,epstein2014,pfeiffer2014,epstein2016hms, iyer2015, chong2016, cole2017, komar2017, sautter2015, hsiao2017, chen2016,staude2013,decker2015}.

These functionalities often rely on the principle of anomalous refraction \cite{YuLightpropagationphase2011}, whereby a gradient of the transmission phase leads to additional refraction of the transmitted beam. This anomalous refraction enables more complex metasurface functionalities, such as lenses, which vary the angle of anomalous refraction across their surface, in order to focus all energy to a point. Therefore, metasurfaces exhibiting uniform anomalous refraction are an important benchmark case. The Huygens' metasurface approach has been quite successful; however, a careful analysis of the electromagnetic boundary conditions \cite{EstakhriWavefrontTransformationGradient2016} has demonstrated that for a passive Huygens' metasurface exhibiting anomalous refraction, some degree of loss or spurious beam generation is unavoidable.

In essence, these imperfections can be understood as arising from imperfect impedance matching. Metasurfaces utilized for anomalous refraction are typically designed to respond only to tangential components of incident fields. For an incident TE polarized wave propagating at angle $\theta$ relative to the surface normal, the ratio of tangential electric to magnetic fields is given by $Z_\mathrm{in} = \eta_0/\cos\theta_\mathrm{in}$, where $\eta_0=\sqrt{\mu_0/\varepsilon_0}$ is the impedance of free space. The transmitted wave also has an impedance $Z_\mathrm{out} = \eta_0/\cos\theta_\mathrm{out}$; however, for anomalous refraction we are specifically interested in the case where $\theta_\mathrm{out}\neq\theta_\mathrm{in}$, hence $Z_\mathrm{out}\neq Z_\mathrm{in}$. Since a Huygens' metasurface is symmetric, it cannot simultaneously be impedance matched to both the incoming and outgoing waves.

It was shown in Refs.~\onlinecite{epstein2016bian,wong2016} that simultaneous impedance matching to the incident and transmitted waves can be achieved with an asymmetric metasurface structure. In contrast to the Huygens' metasurface, such asymmetric structures cannot be fully characterized by their electric and magnetic impedances or susceptibilities. An additional Omega bianisotropy parameter must be introduced, corresponding to the coupling between electric and magnetic responses. The resulting metasurfaces are known as Omega bianisotropic metasurfaces (O-BMS), and can be understood as a generalization of the Huygens' metasurface approach. It was shown in Ref.~\onlinecite{EpsteinHuygensmetasurfacesequivalence2016}, that the theoretically achievable efficiency for the Huygens' metasurface decreases drastically for refraction angles above 55$^\circ$; whereas, the O-BMS can theoretically achieve full efficiency for all angles of refraction. It is important to note that the efficiency refers to the fraction of incident energy which is transmitted at the desired angle. A portion of the remaining energy will be transmitted at unwanted angles, thus in applications such as lenses it may degrade imaging performance.

The advantages of O-BMSs were demonstrated within the framework of the generalized sheet transition conditions \cite{holloway2012}, where the metasurface is modeled by impedance functions which vary continuously with position. However, in practice, this continuous function must be implemented as a discrete array of metamaterial elements. Furthermore, the response of these individual elements is typically characterized within an infinite lattice of identical elements. Due to coupling effects, the response may shift when placed in a super-lattice of non-identical elements. These two effects lead to additional performance degradation, which is not captured within the idealized analytical models. Thus it remains an open question as to whether or not O-BMSs show significantly enhanced performance in practice. To answer this question we compare the efficiency of these two metasurfaces (HMS and O-BMS) for two different refraction angles, 55$^\circ$ and 70$^\circ$.

\section{Structure Design}

\begin{figure}[!ht]
	\centering
	\includegraphics[width=\columnwidth]{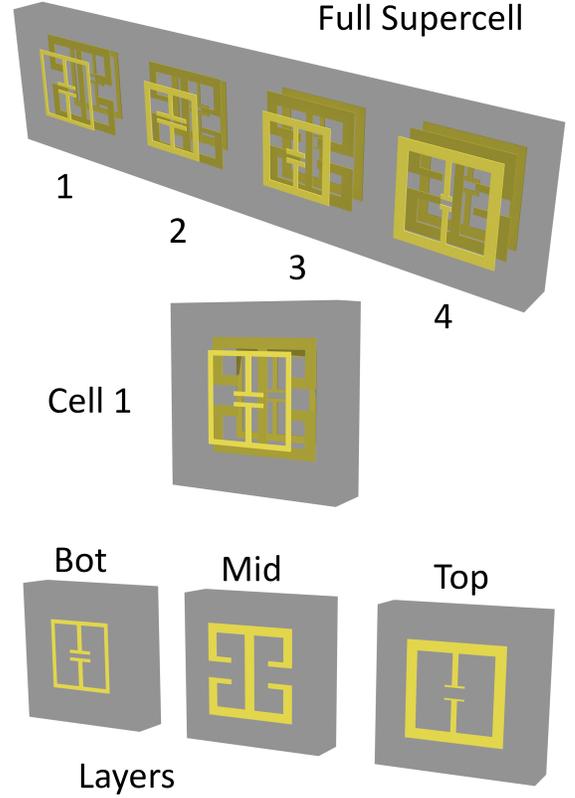}
	\caption[Structure Schematic]{\label{structureschematic} Artistic rendering of the full supercell of the metasurface (top), the individual cell (middle), and the composite layers (bottom) of the first unit cell of the supercell, bot, mid, and top.}
\end{figure}

A supercell of the proposed O-BMS can be seen in the top of Fig.~\ref{structureschematic}. It consists of 3 layers of meta-atoms, separated by dielectric substrates. The structure of the Huygens' metasurface is based on similarly designed patterns, except that it is constrained to have identical top and bottom layers. Each cell in Fig.~\ref{structureschematic} is designed separately, to generate the required electric, magnetic and magneto-electric response at the corresponding position.  

A design procedure for such 3 layer structures was presented in Refs.~\onlinecite{epstein2016bian,pfeiffer2013,pfeiffer2013hms,pfeiffer2014,monticone2013,wong2016}. From the designed angle of incidence and refraction, the phase of the transmission response is given as a linear function with 2$\pi$ phase coverage as
\begin{equation} \label{eq:phase_gradient}
\phi(y) = -ky(\sin\theta_{out} - \sin\theta_{in}).
\end{equation}
%
%Once these values are reached for each layer, the layers are combined to produce each of the four individual cells and the admittance values of each of the four cells are calculated using the following equations from Ref.~\onlinecite{epstein2016bian} 

This transmission phase response can be achieved by a metasurface having the following equivalent magnetic surface admittance $Y_{sm}$ and electric surface impedance $Z_{se}$\cite{epstein2016bian}
\begin{equation}\label{ycell}
Y_{sm} = -j\frac{\overline{Y}_G}{2} \frac{\sin(ky\Delta_{\sin} + \xi_{out})}{1 - (\overline{Z}_A/\overline{Z}_G)\cos(ky\Delta_{\sin} + \xi_{out})}
\end{equation}
\begin{equation}\label{zcell}
Z_{se} = -j\frac{\overline{Z}_G}{2} \frac{\sin(ky\Delta_{\sin} + \xi_{out})}{1 - (\overline{Z}_A/\overline{Z}_G)\cos(ky\Delta_{\sin} + \xi_{out})}.
\end{equation}

In the O-BMS case there is the addition of the Omega-bianisotropy term $K_{em}$, given by
\begin{equation}\label{kcell}
K_{em} = \frac{\Delta Z}{4\overline{Z}_G} \frac{\cos(ky\Delta_{\sin} + \xi_{out})}{1 - (\overline{Z}_A/\overline{Z}_G)\cos(ky\Delta_{\sin} + \xi_{out})}.
\end{equation}
Here $y$ is the cell position within the supercell, $k$ is the wavenumber, $\Delta_{\sin} = \sin\theta_{out} - \sin\theta_{in}$, and $\xi_{out}$, set to zero for this work, is an arbitrary phase term that can be used as another degree of freedom to achieve the desired admittances of the cells. The reamining terms are $\Delta Z = Z_{out} - Z_{in}$, $\overline{Z}_A = (Z_{out} + Z_{in})/2$, $\overline{Z}_G = (Z_{out}Z_{in})^{1/2}$, where $Z_{in} = \eta/\cos\theta_{in}$ and $Z_{out} = \eta/\cos\theta_{out}$ for the transverse electric (TE) polarization we consider. Note that $Y_{sm}$ and $Z_{se}$ are purely imaginary; whereas, $K_{em}$ is purely real. In the HMS case, the structure is matched only to the impedance of the transmitted wave $Z_{out}$, and the term $K_{em}$ is zero, leading to the normalized impedances being equal $Y_{sm}\eta_0=Z_{se}/\eta_0$.

\begin{figure}[!ht]
	\centering
	\includegraphics[width=\columnwidth]{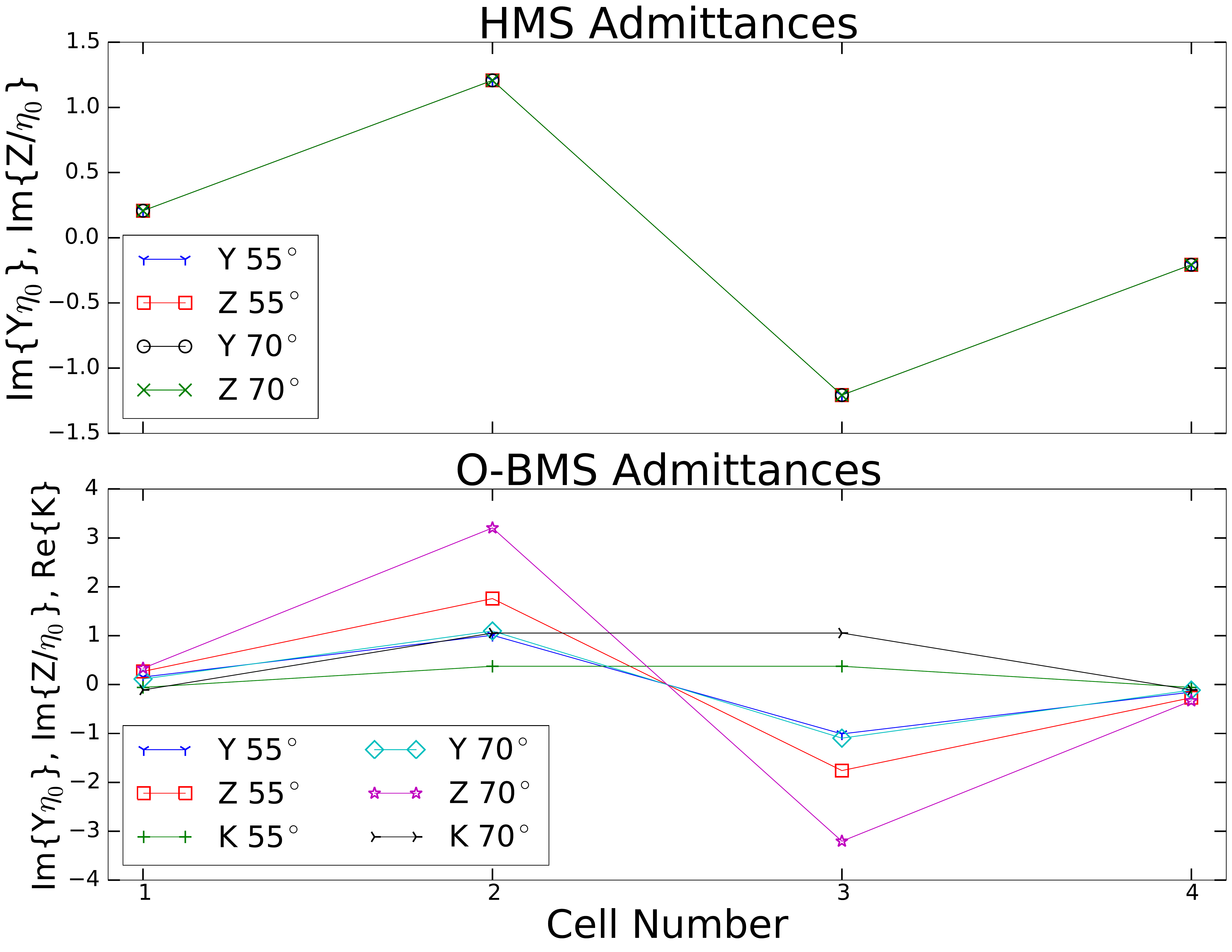}
	\caption[Admittance/Impedance values]{\label{admimpval} The top plot shows the admittance and impedance values as calculated by Eqs.~\eqref{ycell} and \eqref{zcell} for the HMS case for each of the cells and refraction angle. As expected, the admittance and impedance values coincide as the Huygens' condition demands. The lower plot shows the $Y_{sm}$, $Z_{se}$, and $K_{em}$ values for the O-BMS case for each refraction angle. The Omega-bianisotropy term $K_{em}$ must be added in order to achieve impedance matching across the metasurface.}
\end{figure}

We design the structure to operate at 1\,THz, utilizing four cells per supercell, which requires a 90$^\circ$ phase difference between the transmission response of each of the cells. The number of cells is largely determined by fabrication constraints, since meta-atoms with large dimensions are more tolerant to errors in fabrication, but a higher number of cells better approximates the continuous impedance functions specified by the theory. 
For a refraction angle of $55^\circ$, choosing four cells per super-cell leads to a cell size of $\sim$91\,$\mu$m, whereas for $70^\circ$ refraction angle it leads to a cell size of $\sim$80\,$\mu$m. The equivalent impedance functions of each cell are plotted in Fig.~\ref{admimpval}. The relationship between parameters $Y_{sm}$, $Z_{se}$, and $K_{em}$ and the S-parameters of a unit cell obtained from numerical simulation are outlined in the Section I of the Supplemental Material \cite{supplemental}. 

\begin{figure}[!ht]
	\centering
	\includegraphics[width=\columnwidth]{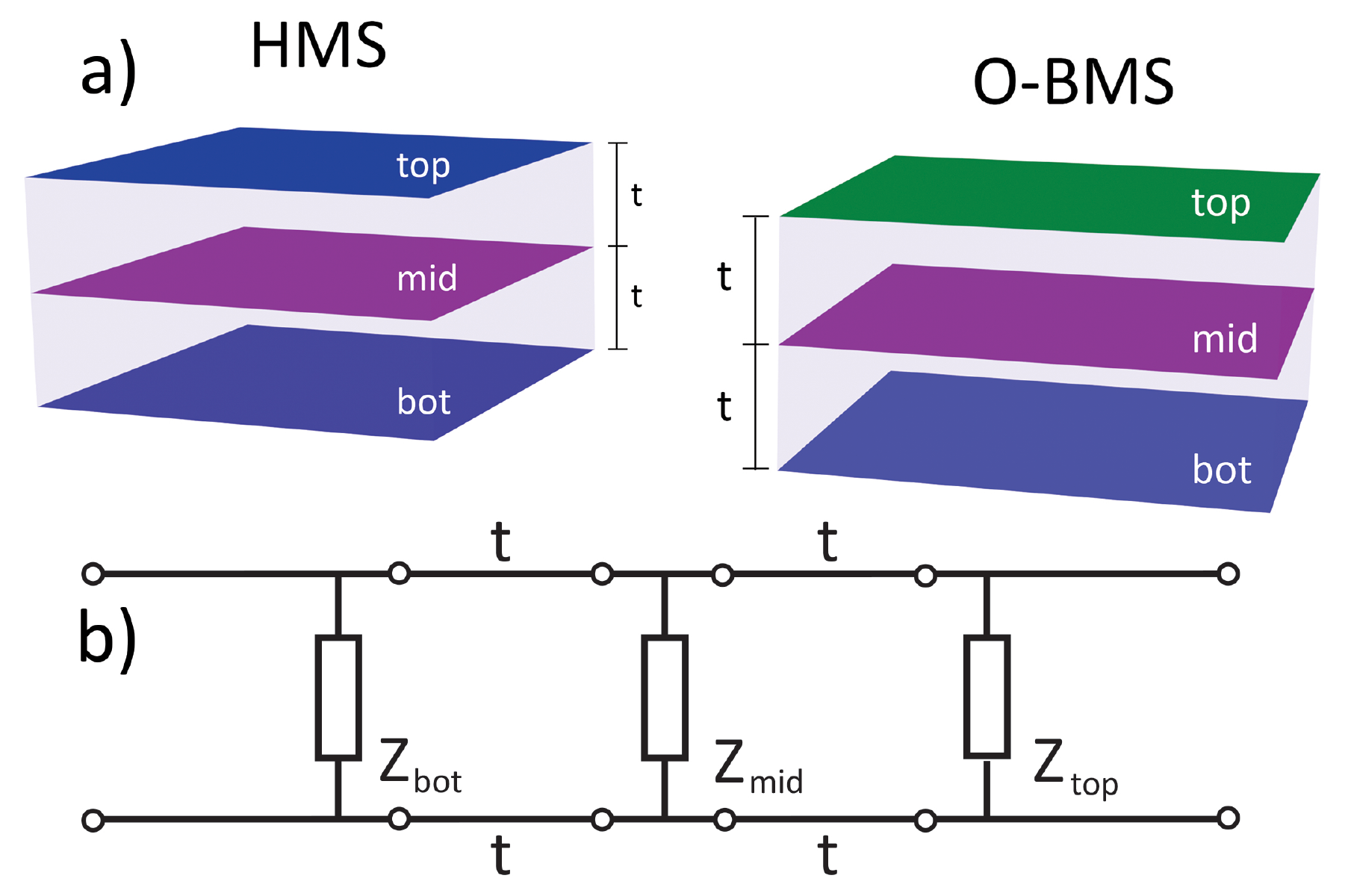}
	\caption[Impedance Sheet Model]{\label{impedsheetmodel} a) Depiction of impedance sheet model for the HMS (left) and O-BMS (right) cases that are compared in this work. HMS case has identical top and bottom layers, while the O-BMS case has different top and bottom layers creating a bianisotropic structure. b) The circuit model representation of the impedance sheets used to model the separate layers of the system.}
\end{figure}

In the approach of Refs.~\onlinecite{epstein2016bian,pfeiffer2014,monticone2013,wong2016}, the effective impedance functions of each cell are realized by a cascade of 3 metallic layers, each represented by a sheet impedance $Z$ or admittance $Y = 1/Z$. 
[See Fig.~\ref{impedsheetmodel}(a)]. For the HMS case the top and bottom sheets are identical but these are different in the O-BMS case. 
 The dielectric substrate between the metallic layers is represented by a transmission line with optical path length $\beta_{sub} t$,  where $\beta_{sub} = \omega \sqrt{\mu_{sub} \epsilon_{sub}}$.  The substrate permittivity $\epsilon_{sub}$ and thickness $t$ are chosen based on fabrication constraints. In Ref.~\onlinecite{epstein2016bian} it was noted that the value $\beta t$ should be small to limit spatial dispersion, which leads to a spurious dependence of the metasurface properties on the angle of incidence.

 As seen in Fig.~\ref{impedsheetmodel}(b), the design method allows each of the three metallic layers to be designed and verified separately, then assembled into the complete cell. In principle, this greatly reduces the complexity of the design process, since it is relatively straightforward to design a single metallic layer to match a given impedance value. However, the transmission-line model, which represents coupling between the metallic layers within one cell, includes only the influence of the fundamental Floquet harmonic, neglecting near-field interaction. We note that this interaction has been shown to be quite strong and to have complex dependence on the distance between meta-atoms \cite{powell2010}. We show subsequently that neglecting this near-field interaction can be a significant source of error in the design process.

To design the individual layers, the required admittance for each layer is calculated using Eq.~(8) in Ref.~\onlinecite{epstein2016bian}. For the HMS case the top and bottom layers of the metasurface are identical and consist of a perfect electric conductor (PEC) element on a dielectric substrate of thickness $t$, as seen in the bottom row of Fig.~\ref{structureschematic}. The middle layer in both cases consists of a PEC element embedded between the two substrate layers. The permittivity $\epsilon$ of the substrate was chosen to be 11.7, corresponding to silicon in the terahertz range (noting that the synthesis procedure requires a lossless substrate material). All simulations had periodic boundary conditions with mesh adaptation enabled. The S-parameter error threshold accuracy was set to 0.0001 with between 60,000 and 80,000 tetrahedrons per layer structure simulation. The geometry of each PEC element is altered until the numerically extracted admittance value equals the theoretical value. The procedure used to convert the S-parameters for each layer to an admittance value is given in Section I the Supplemental Material \cite{supplemental}. This process is repeated for the top and middle layers in the HMS case and the top, middle, and bottom layers for the O-BMS case. The designed and realized admittance values of each layer, for each case of refraction angle, are tabulated in Section II of the Supplemental Material \cite{supplemental}. 

\section{Results for Three-layer Structures}

The three designed layers are assembled into cells and transmission through each cell is simulated with periodic boundary conditions. To take into account the refracted wave impedance, the generalized scattering parameters or G-parameters \cite{monticone2013} are calculated. In the circuit model in Fig.~\ref{impedsheetmodel}(b), this can be understood as connecting different reference impedances $Z_{in}$ and $Z_{out}$ to the respective ports \cite{epstein2016bian}. Perfect impedance matching at these two ports corresponds to the case where reflection coefficients $G_{11} = G_{22} = 0$, and a well designed structure should approximate these conditions.  Achieving full transmission efficiency with the desired transmission phase corresponds to having $G_{12} = G_{21} = e^{i\phi(y)}$, with the phase $\phi(y)$ given by Eq.~\eqref{eq:phase_gradient}. In numerical simulation of a single unit cell, the angle of the transmitted wave will always be identical to the angle of the incident wave, thus it is not possible to directly calculate the generalized scattering parameters. However, they can be obtained by transforming the reference impedances of the numerically obtained S-parameters.

\begin{figure}[!ht]
	\centering
	\includegraphics[width=\columnwidth]{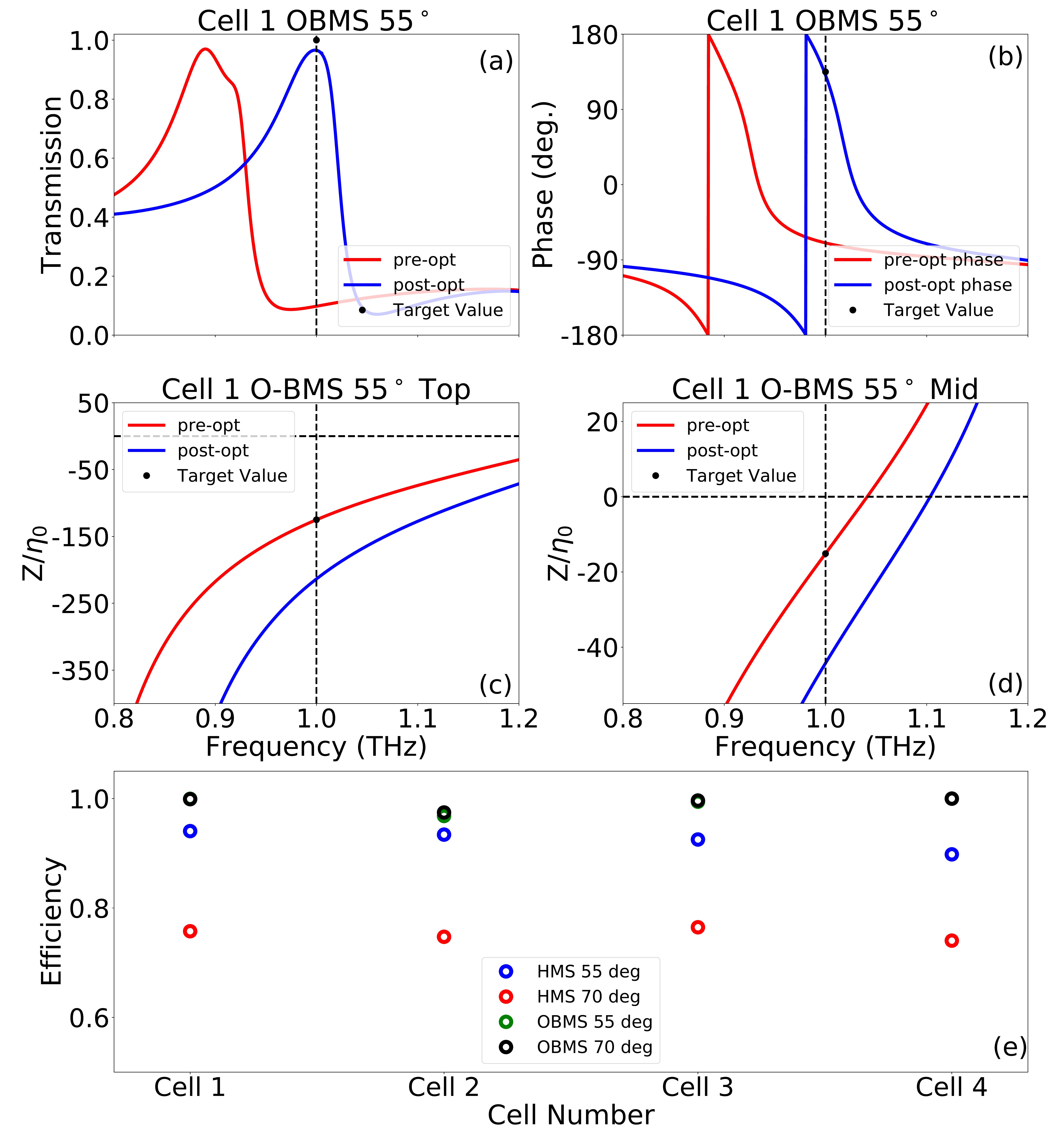}
	\caption[Individual Cell Shift]{\label{fig:cellshift} (a) Magnitude and (b) phase of transmission $|G_{12}|$ of cell 1 of the O-BMS metasurface refracting at 55$^\circ$. The black dots indicate the target values, the red curves indicate the initial results of the synthesis method, and the blue curves indicate the results after numerical optimization. The correspondinding impedances of the (c) top and (d) middle metallic layers. (e) Efficiency $|G_{12}|^2$ of optimized cells at the operating frequency, for all four designed metsaurfaces.}
\end{figure}

The red curves in Fig.~\ref{fig:cellshift}(a) and (b) show the generalized scattering parameters for cell 1 of the O-BMS case for a 55$^\circ$ angle of refraction. It is clear that the transmission magnitude is not equal to the designed value of 1.0 (black dot) at the target frequency of 1\,THz. Instead there is a transmission maximum at a frequency of 0.89 THz. There is a similar shift in the phase of the transmitted wave, where the target value of 135$^\circ$ is indicated by the black dot. This frequency shift is caused by near-field coupling between the layers, which is not accounted for in the transmission-line model. To compensate for this unwanted near-field interaction, the geometric parameters of each of the layer cells are numerically optimized in order to achieve the desired transmission and phase of each cell across the metasurface. The resulting optimization of $|G_{12}|$ for Cell 1 at 55$^\circ$ is shown by the blue curve in Fig.~\ref{fig:cellshift}(a), with the corresponding phase shown by the blue curve in panel (b), nearly reaching the target with value of 133.5$^\circ$.

To gain some insight into the near-field coupling effects which degrade metasurface performance, in Fig.~\ref{fig:cellshift}(c) and (d) we plot the electrical impedance of the top and middle layers of cell 1 of the $55^\circ$ O-BMS structure before and after optimization. These impedance calculations are performed without any other layers present, thus they represent the true self-impedance of each metallic layer. In both cases, we see that the designed impedance value indicated by the black dots is negative, corresponding to a capacitive impedance $Z\sim (j\omega C)^{-1}$. Both optimized structures have a more negative impedance value, corresponding to a reduction in the self-capacitance. This suggests that for these particular layers, mutual capacitance dominates the coupling, leading to an effective increase in the self-capacitance of each layer. The numerical optimization compensates for this by finding a more optimal geometry with lower self capacitance.

Section III of the Supplemental Material\cite{supplemental} shows the impedance values for every layer, many of which follow a similar pattern where the optimized structure has a lower self capacitance. Some of the layers have positive (inductive) impedance $Z\sim j\omega L$. In these cases the optimized values tend to have an increased impedance, corresponding to increased self-inductance. This implies that in such cases coupling is dominated by mutual inductance, which is negative for parallel conductors of the orientation shown in Fig.~\ref{structureschematic}. In general the near-field coupling is complex, and always involves a mix of capacitive and inductive effects\cite{powell2010}, thus not every layer fits the simple patterns described above. When interpreting impedance data it is also important to note that resonances correspond to zeros. The peaks visible in some impedance plots correspond to anti-resonances, where the structure has no influence on the incident field.

%To achieve the Huygens' condition for each HMS cell in numerical simulation, $Y_{sm}\eta_0 = Z_{se}/\eta_0$ is enforced \cite{pfeiffer2013hms}, where the theoretical $Y_{sm}$ and $Z_{se}$ values are calculated according to Eqs.~\eqref{ycell} and \eqref{zcell}. The goals for the O-BMS case are more complicated because the magneto-electric coupling term $K_{em}$ is included to achieve impedance matching. All of these terms are theoretically calculated according to Eqs.~\eqref{ycell}, \eqref{zcell}, and \eqref{kcell}. In order to achieve the proper transmission magnitude and phase of the bianisotropic cell, the goal values in the CST simulations are set equal to the theoretical values. 

\begin{figure}[!th]
	\centering
	\includegraphics[width=\columnwidth]{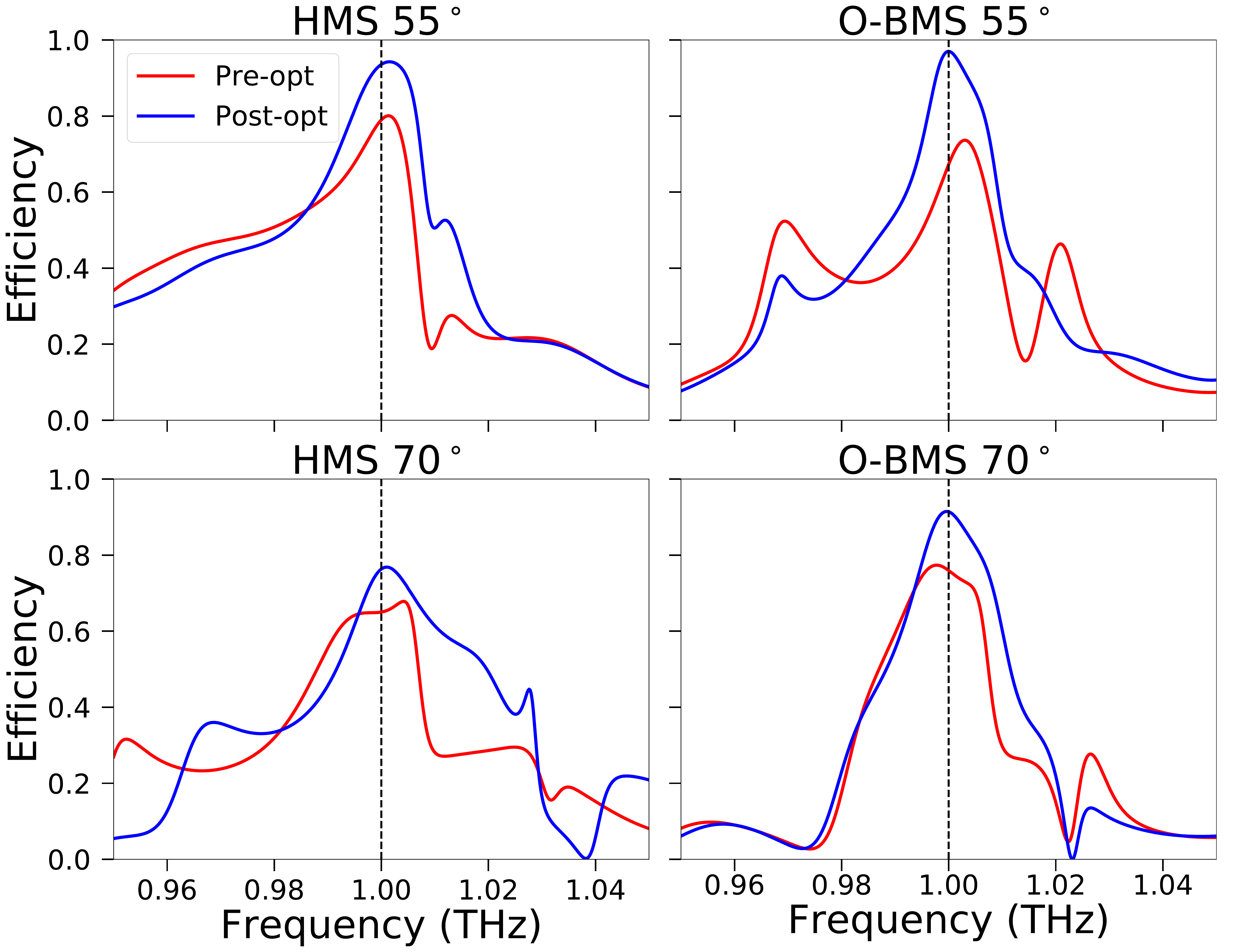}
	\caption[Efficiencies]{\label{efficiency} Red curves show the efficiencies for the full metasurfaces before optimizing the full structure, while the blue curves show the efficiencies of the full metasurface after optimization for both types of metasurface at each refraction angle. The black dashed line indicates the operational frequency of 1.0 THz and the O-BMS case at 70$^\circ$ shows the highest efficiency.}
\end{figure}

The efficiency of each numerically optimized cell $|G_{12}|^2$ is plotted in Fig.~\ref{fig:cellshift}(e). These efficiencies show that the bianisotropic cells (black and green circles) have a higher efficiency than the Huygens' cells (red and blue circles). Furthermore, there is a significantly lower efficiency in the cells of the Huygens' metasurface when the angle is increased from 55$^\circ$ to 70$^\circ$. These results are consistent with theoretical predictions of the reduced efficiency of Huygens' metasurfaces at large refraction angles, in contrast to the robustness of bianisotropic designs \cite{epstein2016bian,wong2016}.

After optimizing each cell individually, they are combined into the supercells in order to create the refracting metasurface. The frequency dependent refraction efficiency of each metasurface is shown by the red curves in Fig.~\ref{efficiency}.  The simulations for each supercell were calculated using mesh adaptation containing approximately 1.2 million tetrahedrons. We note that the efficiency of these supercells is generally lower than that of the individual cells plotted in Fig.~\ref{fig:cellshift}(e). Furthermore, there is some evidence that the optimal frequency is shifted away from the designed frequency. We attribute this to the coupling between neighboring cells, which is not fully accounted for in the design process. In simulating each individual cell, periodic boundary conditions were used, which fully accounts for coupling between identical neighbors. However, once the cells are placed next to non-identical neighbors, the coupling coefficient will change, and the electromagnetic response of each cell will shift from its designed value. To mitigate this frequency shift, we perform a numerical optimization of the entire supercell. The efficiency of the optimized cells is shown by the blue curves in Fig.~\ref{efficiency}, which is significantly improved in all cases.

\begin{figure}[!ht]
	\centering
	\includegraphics[width=\columnwidth]{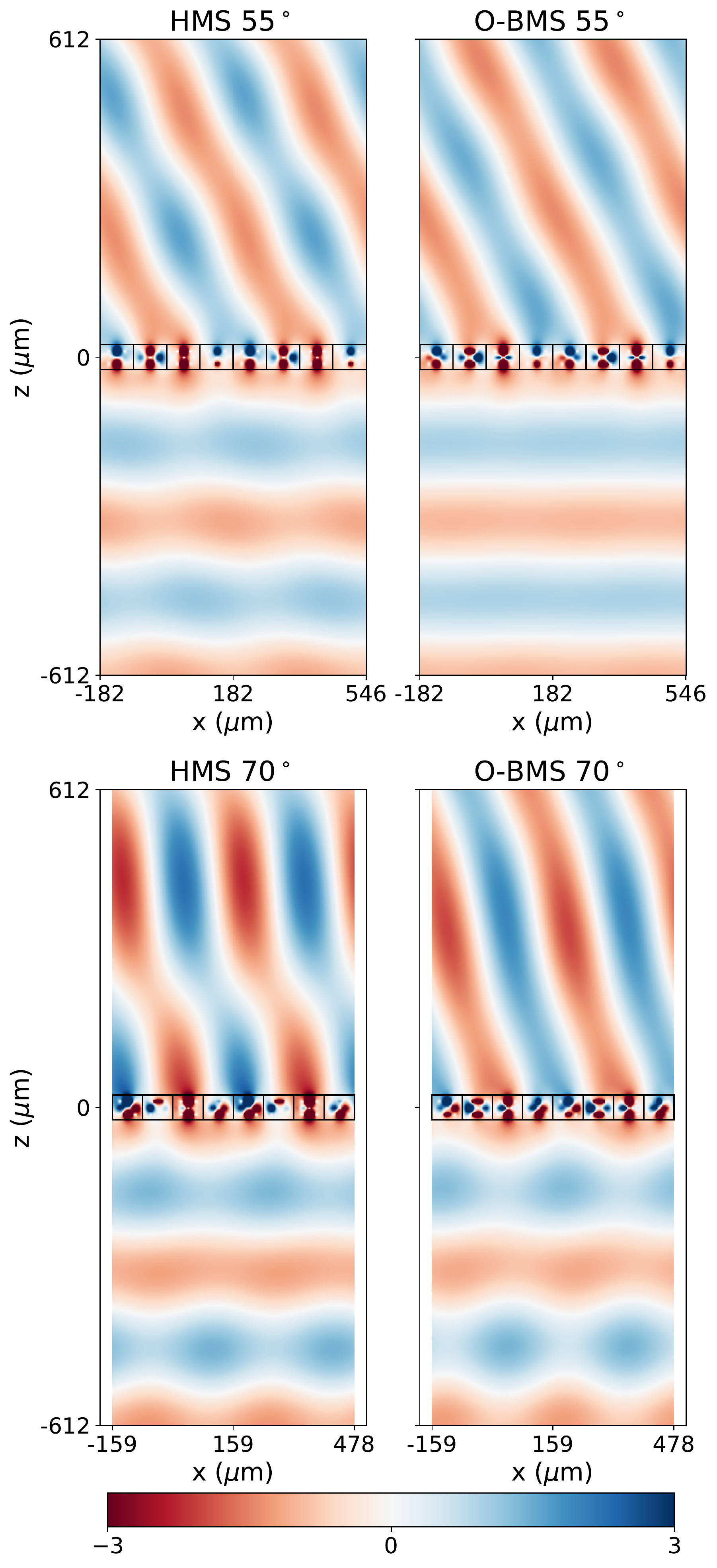}
	\caption[Field Plots]{\label{fieldplots} Electric field plots of the wave scattering for each of the simulated cases. The HMSs are in the left column and the O-BMSs are in the right with the 55$^\circ$ in the top row and the 70$^\circ$ in the bottom.All fields have been normalized to the incident field.}
\end{figure}

Figure \ref{fieldplots} shows the $y$ component of electric field for each optimized structure at the operating frequency, normalized to the incident field amplitude. The wave propagates in the positive $z$ direction, and is refracted in the $x-z$ plane by a metasurface located about $z=0$. The black squares indicate the regions containing each of the unit cells. As expected, Fig.~\ref{efficiency} shows that when the refraction angle is increased from 55$^\circ$ to 70$^\circ$ the HMS efficiency decreases from 93.5\% to 76.4\%. Examining the corresponding field plots in Fig.~\ref{fieldplots}, we see an increase the the strength of the standing wave patterns, corresponding to stronger reflection and transmission of spurious diffraction orders.

For the bianisotropic metasurface refracting at 55$^\circ$, the percentage of power refracted into the desired mode is 94.0\%. This is consistent with the electric field plot in Fig.~\ref{fieldplots} where there is a slight decrease in the unwanted reflections relative to the HMS case. When the angle is increased to 70$^\circ$ the efficiency only decreases slightly to 91.4\%, compared to the 55$^\circ$ case. These results confirm that for higher refraction angles, bianisotropic metasurfaces do offer improved performance.

\section{Conclusion}

In this work we have numerically compared the performance of concrete Huygens' and bianisotropic metasurface designs, in order to verify whether the inclusion of bianisotropy does lead to the efficiency improvements predicted by theory. We utilized an existing design process that breaks the structure into its constituent parts from supercell to cell to layer. We showed that because this design procedure neglects near-field interaction, it does not accurately predict the response when the constituent elements are combined to form the cell and the supercell. This problem can be overcome through a numerical optimization; however, this is computationally expensive when applied to the entire supercell.

We demonstrate that the introduction of bianisotropy into a refracting metasurface can further suppress reflection, creating a highly efficient metasurface in the terahertz regime. We found that the HMS had an efficiency of 93.5\% at an angle of 55$^\circ$ while the O-BMS metasurface was 94.0\%. Once the angle was increased to 70$^\circ$ the efficiency of the HMS plummeted to 76.4\%\% while the O-BMS efficiency remained steady at 91.4\%. This suggests that when considering realistic metasurfaces, the more complex bianisotropic designs are only warranted where large angles of refraction are needed. 

\section*{acknowledgments}
Ilya V. Shadrivov and David A. Powell acknowledge support from the Australian Research Council. 
\section*{References}
\bibliography{ref_paper_bib}

\end{document}

\title{Supplementary Material for Refraction efficiency of Huygens' and bianisotropic terahertz metasurfaces}
\author{Michael A. Cole}
\affiliation{Nonlinear Physics Centre, Research School of Physics and Engineering, The Australian National University, Canberra ACT 2601, Australia}
\email{michael.cole@anu.edu.au}	
\author{Aristeidis Lamprianidis}
\affiliation{Nonlinear Physics Centre, Research School of Physics and Engineering, The Australian National University, Canberra ACT 2601, Australia}
\affiliation{Department of Mathematics and Applied Mathematics, University of Crete, 70013 Heraklion, Crete, Greece}
\author{Ilya V. Shadrivov}
\affiliation{Nonlinear Physics Centre, Research School of Physics and Engineering, The Australian National University, Canberra ACT 2601, Australia}
\author{David A. Powell}
\affiliation{School of Engineering and Information Technology, University of New South Wales, Canberra, ACT 2610, Australia}
\affiliation{Nonlinear Physics Centre, Research School of Physics and Engineering, The Australian National University, Canberra ACT 2601, Australia}

\maketitle

\section{Calculation of equivalent surface properties}\label{sec:extraction}

\subsection{Layer Extraction}

Once the theoretical admittance values for each layer are calculated they become the goal admittances for the CST layer simulations. However, in order to compare the theoretical values to the CST results the element must be de-embedded from the substrate to ensure the correctness of the values. The CST software readily yields the S-parameters; therefore, to obtain the proper values for the admittance of the metallic element, the cascaded sheet admittances presented in References~\onlinecite{epstein2016bian,monticone2013,wong2016,pfeiffer2013hms} are used to determine the simulated admittances. The work in these references describes the use of the \textbf{ABCD} transmission matrix, which relates the field on either side of the cascaded sheet. For example, if the full cell is considered, the \textbf{ABCD} matrix is given by 

\begin{equation}
	\begin{aligned}
		\begin{bmatrix}
		    A       & B  \\
		    C       & D
		\end{bmatrix}
		=
		\begin{bmatrix}
		    1       & 0  \\
		    Y_{e1}       & 1  
		\end{bmatrix}
		\begin{bmatrix}
		    \cos(\beta t)       & j \eta_t \sin(\beta t)  \\
		    j \sin(\beta t)/\eta_t       & \cos(\beta t)  
		\end{bmatrix}\\
		\begin{bmatrix}
		    1       & 0  \\
		    Y_{e2}       & 1  
		\end{bmatrix}
		\begin{bmatrix}
		    \cos(\beta t)       & j \eta_t \sin(\beta t)  \\
		    j \sin(\beta t)/\eta_t       & \cos(\beta t)  
		\end{bmatrix}
		\begin{bmatrix}
		    1       & 0  \\
		    Y_{e3}       & 1  
		\end{bmatrix}.
	\end{aligned}
\end{equation}

The matrices  
\[
\begin{bmatrix}
    1       & 0  \\
    Y_{en}       & 1  \\
\end{bmatrix}
\]
are the transfer matrices (\textbf{ABCD}) of the nth metallic element. The \textbf{ABCD} matrices separating the elements 
\[
\begin{bmatrix}
    \cos(\beta t)       & j \eta_t \sin(\beta t)  \\
    j \sin(\beta t)/\eta_t       & \cos(\beta t)  \\
\end{bmatrix}
\]
represent the fields on either side of the substrate layers, where $t$ is the thickness of the substrate layer and $\beta$ is the wave number in the dielectric. 

Subsequently, to de-embed the metallic element from the substrate the \textbf{ABCD} matrix equation needs to be solved for the admittance $Y_{en}$ in terms of S-parameters. Re-arranging the \textbf{ABCD} matrix equation and using the transformation equations in Ref.~\onlinecite{pozar2004}. The equations for the admittance of the layer elements become

\begin{equation}\label{cst_top_layer}
	\begin{aligned}
	Y_{top/bot} = -\frac{\cos(\beta t)(S_{11} + S_{22} - S_{11} S_{22} + S_{12}S_{21} - 1)}{2S_{21}\eta_0}\\ 
	- \frac{j\sin(\beta t)(S_{22} - S_{11} - S_{11}S_{22} + S_{12}S_{21} + 1)}{2S_{21}\eta_t}
	\end{aligned}
\end{equation}

\begin{equation}\label{cst_mid_layer}
	\begin{aligned}
	Y_{mid} = -\cos(\beta t)\left[\frac{\cos(\beta t)(S_{12}S_{21} - (S_{11} - 1)(S_{22} - 1))}{2S_{21}\eta_0}\right.\\ 
	\left. + \frac{j\sin(\beta t)(S_{12}S_{21} - (S_{11} + 1)(S_{22} - 1))}{2S_{21}\eta_t}\right]\\
	 - j\sin(\beta t)\left[\frac{\cos(\beta t)(S_{12}S_{21} - (S_{11} - 1)(S_{22} + 1))}{2S_{21}\eta_t}\right.\\
	 \left. + \frac{j\eta_0\sin(\beta t)(S_{12}S_{21} - (S_{11} + 1)(S_{22} + 1))}{2S_{21}\eta_t}\right].
	\end{aligned}
\end{equation}

The impedance of free space $\eta_0$ is 376.73 $\Omega$ and $\eta_t$ for the chosen substrate material has a value of 110.14 $\Omega$. At the designed operating frequency $\beta t=0.86$. 

\subsection{Cell Extraction - HMS}
The design process for the individual cells for each of the two cases (HMS and O-BMS) diverge slightly and will be addressed separately. In order to compare the theoretical admittance and impedance values with the simulated results from CST for the HMS case the following equations were used to extract these values from the CST S-parameters according to 

\begin{equation}\label{cst_hms_adm}
Y_{es} = \frac{2(1 - S_{12} - S_{11})}{\eta_0(1 + S_{12} + S_{11})},
\end{equation}

\begin{equation}\label{cst_hms_imp}
Z_{ms} = \frac{2\eta_0(1 - S_{12} - S_{11})}{(1 + S_{12} - S_{11})}
\end{equation}

where $S_{12}$ and $S_{11}$ are the complex transmission and reflection parameters respectively \cite{pfeiffer2013,holloway2005}. This process was conducted for both refraction angles and the cells were optimized so the above extracted numbers from CST matched as closely as possible to the design requirements shown in the top plot of Fig.~2 in the main text. 

\subsection{Cell Extraction - O-BMS}
The added coupling term for the bianisotropic metasurface means that the desired transmission and phase required of the cells will no longer be dictated completely by the admittance and impedance of each cell and the Huygens' condition will no longer be the goal to achieve in the simulations. Furthermore, Equations~\eqref{cst_hms_adm} and \eqref{cst_hms_imp} also no longer hold true. 

The result of this, is that a more complicated post-processing step is needed in order to calculate $Y_{sm}$, $Z_{se}$, and $K_{em}$ based on the S-parameters in CST. The first step began with Equation (7) in Ref.~\onlinecite{epstein2016bian}, which gives the cell parameters in terms of the impedance matrix elements, Z-parameters $Z_{11}$, $Z_{12}$, and $Z_{22}$, as shown in Equations~\eqref{Y},\eqref{Z}, and \eqref{K}, yielding

\begin{equation}\label{K}
K_{em} = \frac{Z_{11} - Z_{22}}{2(Z_{11} + Z_{22} - 2Z_{12})}
\end{equation}

\begin{equation}\label{Y}
Y_{sm} = \frac{1}{Z_{11} + Z_{22} - 2Z_{12}}
\end{equation}
\cite{pozar2004}
\begin{equation}\label{Z}
Z_{se} = \frac{Z_{11}Z_{22} - (Z_{12})^2}{Z_{11} + Z_{22} - 2Z_{12}}.
\end{equation}

Then converting the Z-parameters to S-parameters \cite{pozar2004}, these equations become 

\begin{equation}
K_{em} = \frac{-(S_{11} - S_{22})}{4S_{12} + 2S_{11}S_{22} - 2S_{12}S_{21} - 2}
\end{equation}

\begin{equation}
Y_{sm} = \frac{S_{11} + S_{22} - S_{11}S_{22} + S_{12}S_{21} - 1}{4S_{21} + 2S_{11}S_{22} - 2S_{12}S_{21} - 2}
\end{equation}

\begin{widetext}
\begin{equation}
Z_{se} = \frac{4(S_{12})^2 -(S_{12}S_{21} - (S_{11} - 1)(S_{22} + 1))(S_{12}S_{21} - (S_{11} + 1)(S_{22} - 1))}{(S_{12}S_{21} - (S_{11} - 1)(S_{22} - 1))[(S_{12}S_{21} - (S_{11} - 1)(S_{22} + 1)) + (S_{12}S_{21} - (S_{11} + 1)(S_{22} - 1)) - 4S_{12}]}.
\end{equation}
\end{widetext}

The O-BMS structures were then optimized so that these values matched as closely as possible to the design requirements as shown in the bottom plot in Fig.~2 in the main text.

\section{Admittance of metallic layers}
The values in Table~\ref{table1} are the design target and simulated admittances for the layers of the metasurfaces for each case of refraction angle. 

\begin{table}[htb]
%\begin{tabular}{|p{0.35cm}|p{0.35cm}|p{0.86cm}|p{0.86cm}|p{0.86cm}|p{0.86cm}|p{0.86cm}|p{0.86cm}|p{0.86cm}|p{0.86cm}|}
\begin{tabular}{|p{0.5cm}|p{0.5cm}|p{1cm}|p{1cm}|p{1cm}|p{1cm}|p{1cm}|p{1cm}|p{1cm}|p{1cm}|}
	\hline
	& & \multicolumn{2}{|c|}{HMS 55$^{\circ}$} & \multicolumn{2}{|c|}{HMS 70$^{\circ}$} & \multicolumn{2}{|c|}{O-BMS 55$^{\circ}$} & \multicolumn{2}{|c|}{O-BMS 70$^{\circ}$}\\
	\hline
%	&  & $Y_{th}\eta_0$ & $Y_{sm}\eta_0$ & $Y_{th}\eta_0$ & $Y_{sm}\eta_0$ & $Y_{th}\eta_0$ & $Y_{sm}\eta_0$ & $Y_{th}\eta_0$ & $Y_{sm}\eta_0$ \\
	&  & Des & Sim & Des & Sim & Des & Sim & Des & Sim \\
		\hline
	\parbox[t]{2mm}{\multirow{3}{*}{\rotatebox[origin=c]{90}{Cell 1}}} & T & 0.0089
	& 0.0089 & 0.0089 & 0.0089 & 0.0080 & 0.0080 & 0.0074 & 0.0074 \\
	& M & 0.0538 & 0.0539 & 0.0538 & 0.0538 & 0.0661 & 0.0661 & 0.0810 & 0.0811 \\
	& B & 0.0089
	& 0.0089 & 0.0089 & 0.0089 & 0.0091 & 0.0091 & 0.0091 & 0.0091 \\
	\hline
	\parbox[t]{2mm}{\multirow{3}{*}{\rotatebox[origin=c]{90}{Cell 2}}} & T & 0.0142 & 0.0143 & 0.0142 & 0.0143 & 0.0133 & 0.0120 & 0.0127 & 0.0127 \\
	& M & 0.0538 & 0.0539 & 0.0538 & 0.0538 & 0.0661 & 0.0661 & 0.0810 & 0.0811 \\
	& B & 0.0142 & 0.0143 & 0.0142 & 0.0143 & 0.0122 & 0.0123 & 0.0109 & 0.0109 \\
	\hline
	\parbox[t]{2mm}{\multirow{3}{*}{\rotatebox[origin=c]{90}{Cell 3}}} & T & 0.0014 & 0.0014 & 0.0014 & 0.0015 & 0.0023 & 0.0023 & 0.0030 & 0.0030 \\
	& M & -0.023 & -0.023 & -0.023 & -0.023 & -0.035 & -0.036 & -0.050 & -0.050 \\
	& B & 0.0014 & 0.0014 & 0.0014 & 0.0015 & 0.0035 & 0.0035 & 0.0047 & 0.0047 \\
	\hline
	\parbox[t]{2mm}{\multirow{3}{*}{\rotatebox[origin=c]{90}{Cell 4}}} & T & 0.0067 & 0.0067 & 0.0067 & 0.0067 & 0.0076 & 0.0076 & 0.0083 & 0.0083 \\
	& M & -0.023 & -0.023 & -0.023 & -0.023 & -0.035 & -0.036 & -0.050 & -0.050 \\
	& B & 0.0067 & 0.0067 & 0.0067 & 0.0067 & 0.0065 & 0.0065 & 0.0065 & 0.0065 \\
	\hline
\end{tabular}
\caption{\label{table1}
Designed (Des) and simulated (Sim) admittance values of the layers (T-top, M-middle, and B-bottom) for each type of metasurface, for the two cases of refraction angle. All values are normalized as $Y\eta_0$. The admittances for the top and bottom layers of the HMS metasurfaces are identical and they are different for the bianisotropic case.}
\end{table}

\section{Influence of near-field coupling within cells}\label{sec:layer_imp}
The following figures show the impedance of each layer before and after optimization of the assembled cell, as well as the corresponding transmission amplitudes of each complete cell. The red curves represent the pre-optimized values obtained by a direct application of the design approach. The blue curves are the values obtained after performing numerical optimization. 

\begin{figure}[!htb]
	\centering
	\includegraphics[width=\columnwidth]{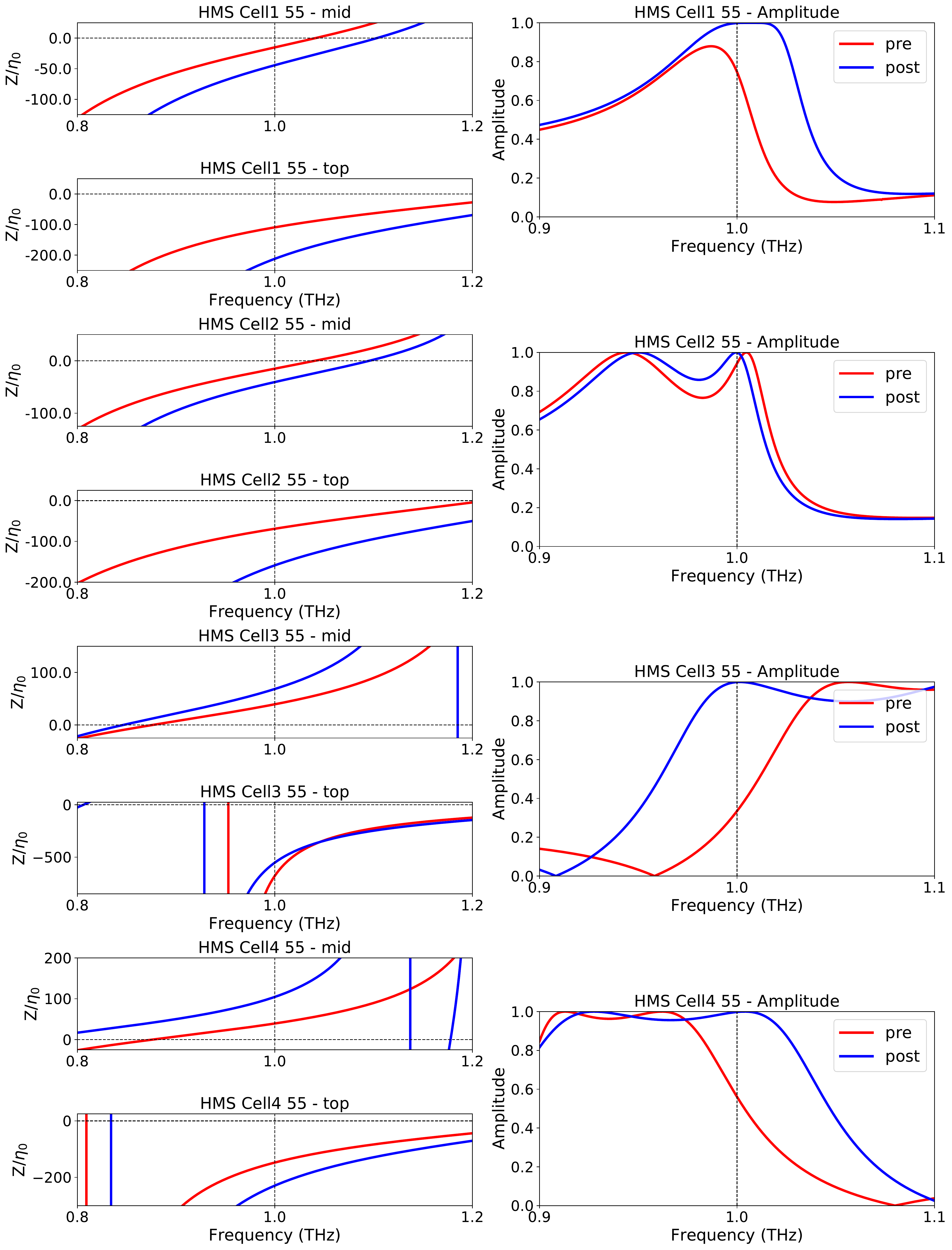}
	\caption[HMS 55 Impedances]{\label{hms55imp} Layer impedance and transmission response before and after optimization of the cells for the HMS designed for a refraction angle of 55$^\circ$.}
\end{figure}

\begin{figure}[!htb]
	\centering
	\includegraphics[width=\columnwidth]{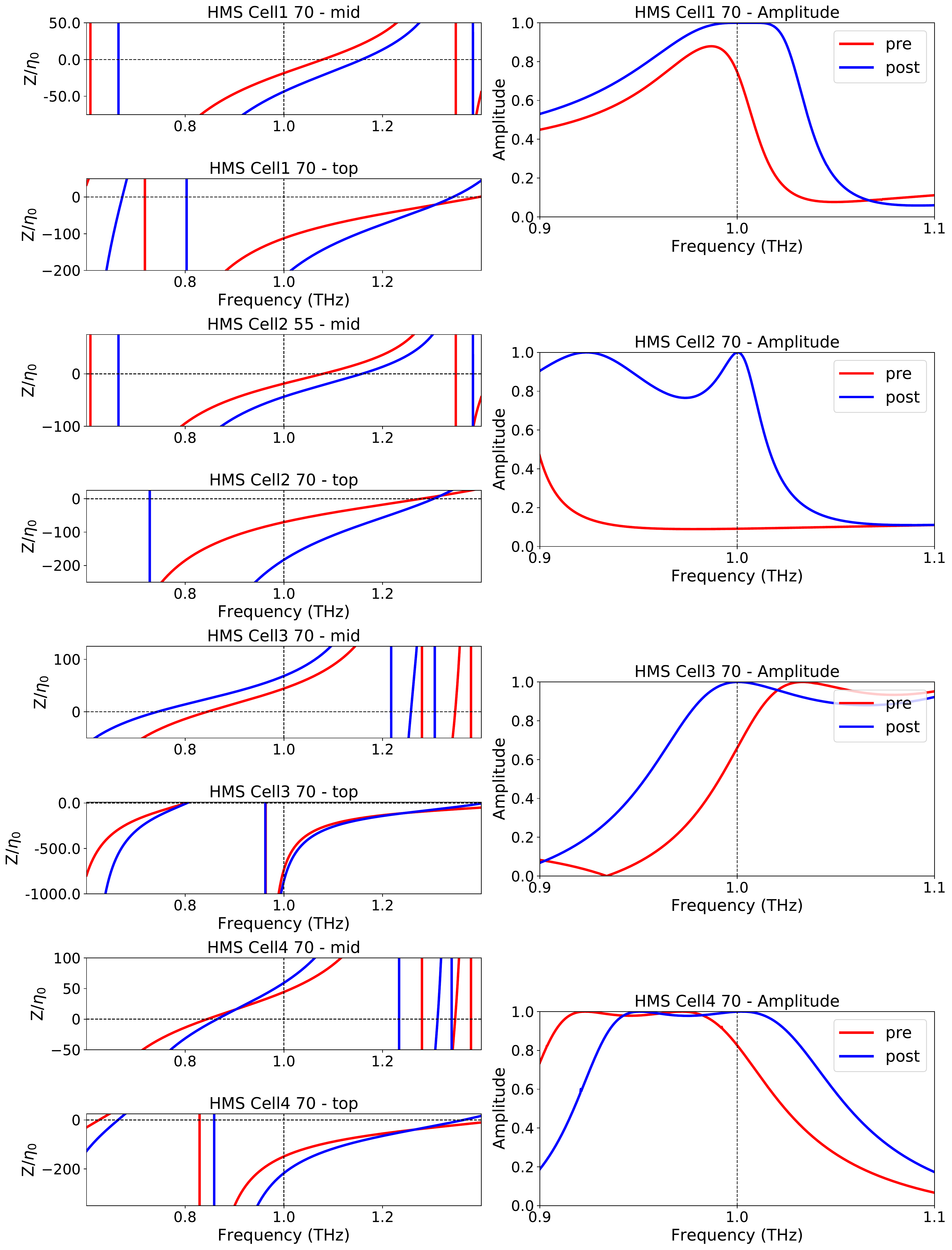}
	\caption[HMS 70 Impedances]{\label{hms70imp} Layer impedance and transmission response before and after optimization of the cells for the HMS designed for a refraction angle of 70$^\circ$.}
\end{figure}

\begin{figure}[!htb]
	\centering
	\includegraphics[width=\columnwidth]{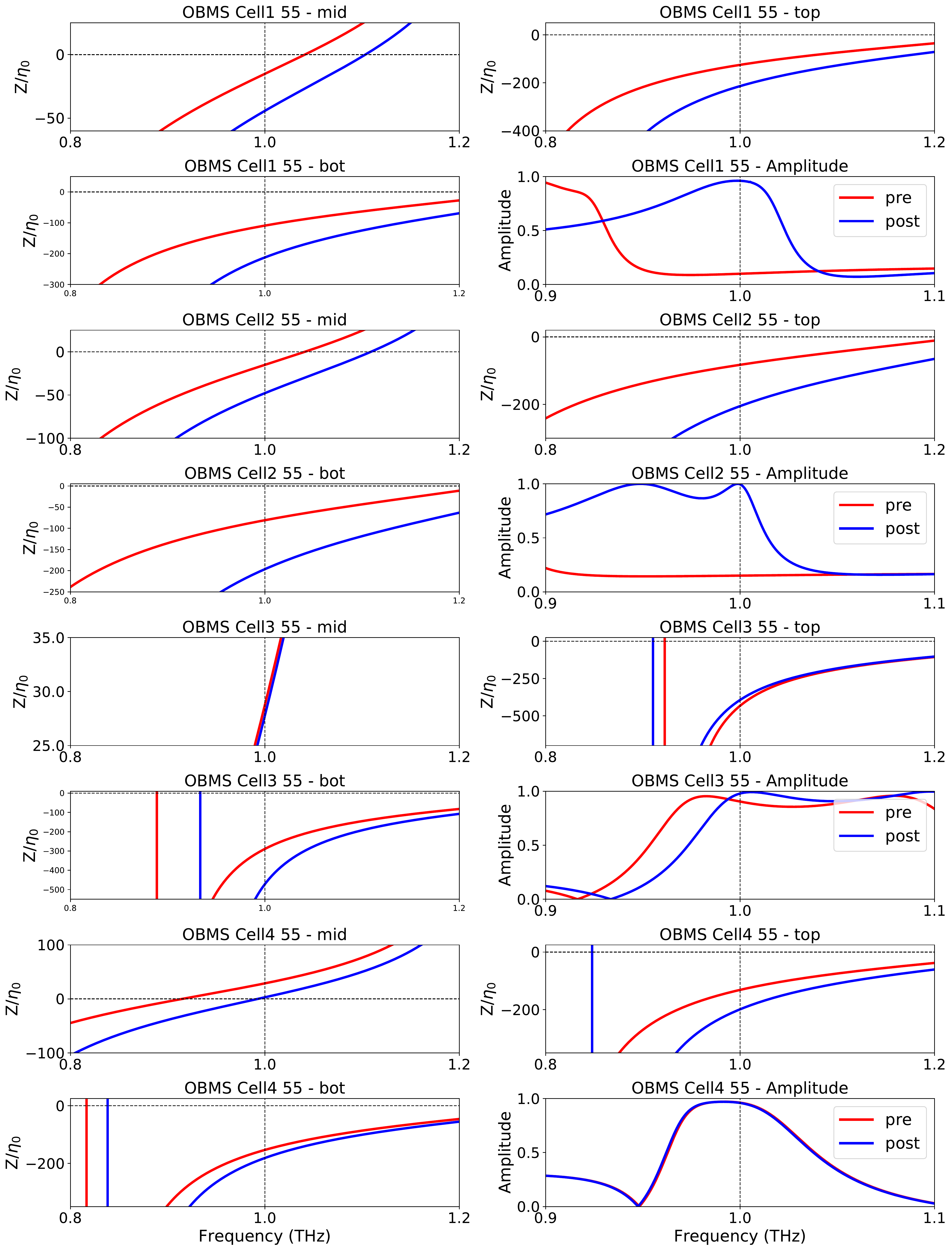}
	\caption[O-BMS 55 Impedances]{\label{obms55imp} Layer impedance and transmission response before and after optimization of the cells for the O-BMS designed for a refraction angle of 55$^\circ$.}
\end{figure}

\begin{figure}[!htb]
	\centering
	\includegraphics[width=\columnwidth]{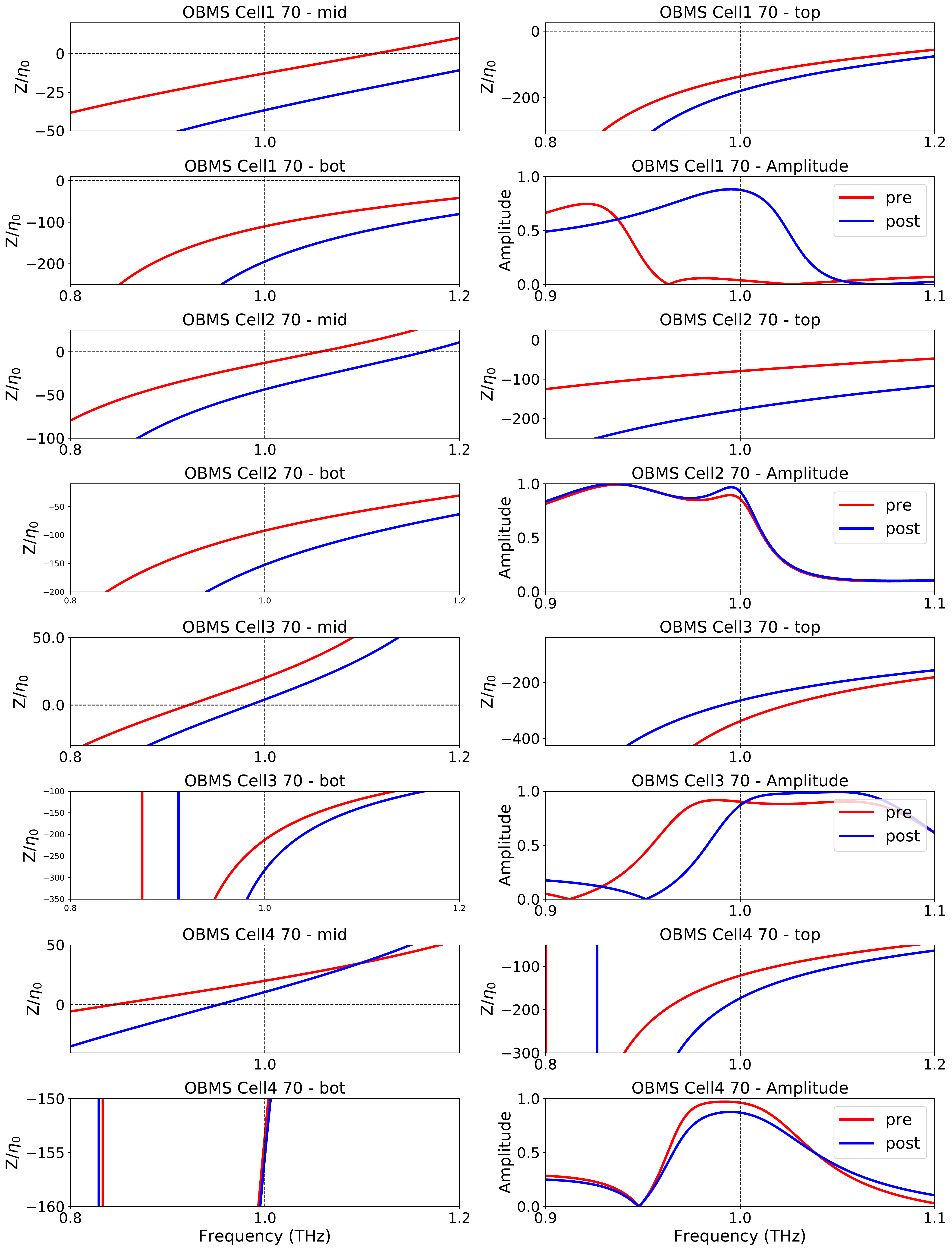}
	\caption[O-BMS 70 Impedances]{\label{obms70imp} Layer impedance and transmission response before and after optimization of the cells for the O-BMS designed for a refraction angle of 70$^\circ$.}
\end{figure}

\clearpage

\bibliography{ref_paper_bib}